\documentclass[aps,twocolumn,showpacs]{revtex4}
\usepackage{graphicx} 
\usepackage{caption}  
\usepackage{subcaption}
\usepackage{amsmath,color,ulem}
\usepackage{physics}
\usepackage{epsfig}
\usepackage{multirow}
\usepackage{float}
\usepackage{array} 
\usepackage{rotating}

\usepackage{booktabs} 

\begin{document}
	
	\title{Mass Spectra and Radiative Decay Widths of \(D_{s0}^{*}(2317)^+\), \(D_{s1}(2460)^+\) and \(D'_{s1}(2536)^+\)}
\author{Shirui Zhao$^1$}
\author{Yuxuan Du$^1$}
\author{Manying Pan$^1$}\email[E-mail: ]{mypan@njnu.edu.cn}
\author{Hongxia Huang$^1$}\email[E-mail: ]{hxhuang@njnu.edu.cn}
\author{Jialun Ping$^1$}\email[E-mail: ]{jlping@njnu.edu.cn}
\affiliation{$^1$Department of Physics, Nanjing Normal University, Nanjing, Jiangsu 210097, China}

  \begin{abstract}
		Inspired by recent experimental reports on the radiative decay \(D_{s1}(2460)^{+}\to D_s^{*+}\gamma\) and the branching-fraction ratio \(\mathcal{B}[D_{s1}(2460)^{+}\to D_s^{*+}\gamma]/\mathcal{B}[D_{s1}(2460)^{+}\to D_s^{+}\gamma]\), we investigate the mass spectra and radiative decay widths of \(D_{s0}^{*}(2317)^+\), \(D_{s1}(2460)^+\) and \(D_{s1}'(2536)^+\).
		Within the chiral quark model combined with the Gaussian Expansion Method, the mass spectra obtained agree reasonably well with the experimental data. 
		Besides, the branching-fraction ratio obtained in this work is \(\mathcal{B}[D_{s1}(2460)^{+}\to D_s^{*+}\gamma]/\mathcal{B}[D_{s1}(2460)^{+}\to D_s^{+}\gamma] = 0.57\) and \(\Gamma(D_{s0}^{*}(2317)^{+}\to D_s^{*+}\gamma) = 3.80\,\mathrm{keV}\). 
		By analyzing our theoretical results against the latest experimental measurements, we conclude that the \(D_{s1}(2460)^{+}\) can be treated as a mixed state of \(1^1P_1\) and \(1^3P_1\), and it may contain both the \(c\bar{s}\) component and the tetraquark component. 
		In addition, we also obtain \(\Gamma(D'_{s1}(2536)^{+}\to D_s^{*+}\gamma) = 9.14\,\mathrm{keV}\) and \(\Gamma(D'_{s1}(2536)^{+}\to D_s^{+}\gamma) = 8.59\,\mathrm{keV}\) by treating \(D'_{s1}(2536)^+\) as a meson state. 
		More precise experimental data are required to pin down the internal structure of these states.
		
	\end{abstract}
	
	\maketitle
	
	\setcounter{totalnumber}{5}
	
  \section{\label{sec:Introduction}Introduction}
	Very recently, the Belle and Belle II Collaborations reported the first observation of the radiative decay $D_{s1}(2460)^+ \to D_s^{*+}\gamma$ with a significance greater than 10 standard deviations, together with a measurement of the branching-fraction ratio $\mathcal{B}[D_{s1}(2460)^+ \to D_s^{*+}\gamma]/\mathcal{B}[D_{s1}(2460)^+ \to D_s^+\gamma] = 0.74 \pm 0.05\,(\text{stat.}) \pm 0.06\,(\text{syst.})$ \cite{draft}.
	Previously, the same Collaborations also reported the radiative decay $D^*_{s0}(2317)^+ \to D_s^{*+}\gamma$, with the branching-fraction ratio $\mathcal{B}(D^*_{s0}(2317)^+ \to D_s^{*+}\gamma)/\mathcal{B}(D^*_{s0}(2317)^+ \to D_s^+\pi^0)$ measured to be $[7.14 \pm 0.70\,(\text{stat.}) \pm 0.23\,(\text{syst.})]\%$ \cite{ootr}.
	In fact, a wealth of experimental studies and reports on these two states already exists.
	As early as 2003, the Belle Collaboration observed \(D^*_{s0}(2317)^+\) and \(D_{s1}(2460)^+\) in exclusive \(B\)-meson decays, with \(D^*_{s0}(2317)^+\) reconstructed via \(D^*_{s0}(2317)^+ \to D_{s}^{+}\pi^{0}\). The radiative decays \(D^*_{s0}(2317)^+ \to D_{s}^{*+}\gamma\) and \(D_{s1}(2460)^+ \to D_{s}^{*+}\gamma\) were both constrained by \(90\%\) C.L. upper limits, while a helicity analysis of \(D_{s1}(2460)^+ \to D_s^+\gamma\) favoured the assignment \(J^P=1^+\) \cite{oot2457}.
	In the following years, continuous experimental efforts by Belle, BaBar, LHCb, and other collaborations have extensively investigated the properties of these two states \cite{sobd,motd,sobd2,asot,mota,sod2460}.
	
    The discoveries of the charm-strange mesons \(D^*_{s0}(2317)^{+}\) and \(D_{s1}(2460)^{+}\) have stimulated a number of theoretical studies. In fact, both states have long attracted attention due to a series of anomalous experimental properties: they are narrow resonances, with dominant strong decays that violate isospin \cite{ooan2317,ooan2460}, and measured masses roughly 100~MeV or more below the predictions of early conventional quark-potential models \cite{miar,tpop,lsah,hmsa}.
    Motivated by these peculiar properties, theorists have proposed a variety of models to describe their underlying structure, such as pure meson states \cite{ttno,cmoh,potc,understanding,rtod,pmca,poec,etot,podm}, molecular states \cite{ioad,ohlm,psbo,uosa,acfi}, compact tetraquarks \cite{bdas,tbbt,daaf,dwho,tqsi,sotn}, and mixed states \cite{cotn,tiif,masf,rolq,hmit,tmsa,mard,ssod}.
    
    For \(D^*_{s0}(2317)^{+}\), within the pure meson picture, the relativistic quark model based on the Godfrey-Isgur framework assigns it as the \(1^3P_0\) state, reproducing its mass at 2317~MeV and computing a radiative width of about 2~keV for \(D^*_{s0}(2317)^{+} \to D_s^{*+}\gamma\) \cite{etot}.
    Alternatively, based on Chiral Effective Field Theory, it is supported as a hadronic molecule dominated by the isoscalar \(DK\) channel, with its mass fixed at 2318~MeV and a corresponding radiative width of 3.7~keV \cite{uosa}. 
    In the compact tetraquark picture, a non-relativistic quark potential model describes it as a diquark-antidiquark bound system. The initial mass calculated with universal parameters is about 150~MeV higher than the experiment, whereas after refitting the interaction parameters using \(D^*_{s0}(2317)^{+}\) and \(D_{s1}(2460)^{+}\) as inputs, the predicted mass aligns with the experimental value \cite{sotn}. 
    Furthermore, some unquenched quark models also treat it as a mixed state of a bare \(c\bar{s}\) core and a \(DK\) component \cite{ssod, mard}. Ref.~\cite{ssod} introduces a damping factor into the \(^3P_0\) operator within the chiral quark model (ChQM) to constrain the mass shift downwards, while Ref.~\cite{mard} incorporates both the \(c\bar{s}-DK\) coupling and the \(DK-DK\) self-interaction, successfully lowering the theoretical mass to match the experimental value. After including the \(D^{(*)}K\) interaction, Ref.~\cite{mard} reports that, for cutoff parameters of 0.5, 1.0, and 1.5~GeV, the calculated radiative decay widths of \(D_{s0}^*(2317) \to D_s^*\gamma\) are 1.41, 4.27, and 3.12~keV, respectively.
    
    For \(D_{s1}(2460)^{+}\), using the relativized quark model, Ref.~\cite{poec} treats it as a mixed state of the \(1^1P_1\) and \(1^3P_1\) configurations for the calculation, yielding a mass about 90~MeV higher than the experiment and a radiative width of 5.4~keV for \(D_{s1}(2460)^{+} \to D_s^{*+}\gamma\).
    On the other hand, within the molecular framework, chiral effective field theory yields a \(D^*K\)-dominated molecule with a self-consistent mass of 2458~MeV, which agrees with the experiment, and a radiative width of 13~keV for \(D_{s1}(2460)^{+} \to D_s^{*+}\gamma\) \cite{uosa}.
    Within the compact tetraquark picture, Ref.~\cite{sotn} identifies \(D_{s1}(2460)^+\) as a diquark-antidiquark S-wave ground state where the constituent \(cq\) diquark and \(\bar{s}\bar{q}\) antidiquark both have an internal spin of 1, and based on their shared \(L=0\) nature, classifies it alongside \(D_{s0}^*(2317)^+\) as a pair of ground states within the same tetraquark multiplet.
    Furthermore, the unquenched mixed‑state picture interprets $D_{s1}(2460)^{+}$ as a physical mixture of a bare $c\bar{s}$ core and $D^*K$ meson‑meson components\cite{mard,ssod}. 
    Within the ChQM, Ref.~\cite{ssod} employs the $^3P_0$ transition operator modified by a damping factor to realize configuration coupling between two‑quark and four‑quark Fock components, yielding a $D^*K$ meson‑meson component fraction of approximately 52.2\% in the physical state.
    By simultaneously including the channel-coupling effects and the \(D^*K\)-\(D^*K\) two-hadron self-interaction, Ref.~\cite{mard} obtains a consistent mass of 2456~MeV. Furthermore, for cutoff parameters of 0.5, 1.0, and 1.5~GeV, the calculated radiative decay widths of \(D_{s1}(2460)^+ \to D_s^{*+}\gamma\) are 1.74, 2.19, and 2.75~keV, and the corresponding widths for \(D_{s1}(2460)^+ \to D_s^+\gamma\) are 0.98, 6.41, and 6.99~keV, respectively.
    In summary, although these pictures can partially explain the available data, a consensus on the internal configurations of both \(D^*_{s0}(2317)^+\) and \(D_{s1}(2460)^+\) has not yet been reached.
   
    In general, many theoretical works fail to simultaneously reproduce the mass spectra and decay widths in a self-consistent manner, and some studies adopt the parameters fitted to experimental mass spectra as inputs for the calculation of decay widths. 
    In fact, to provide a more self-consistent interpretation of their internal configurations, a joint analysis combining both the mass spectra and radiative decay properties is highly desirable. 
    Furthermore, given the recent experimental updates on the radiative decays of these states, such a joint investigation is both timely and necessary. 
    Motivated by this, we systematically investigate \(D^*_{s0}(2317)^+\) and \(D_{s1}(2460)^+\) from the perspectives of both mass spectra and radiative decays in this work.
    
    Besides, within the same \(J^P=1^+\) channel, alongside \(D_{s1}(2460)^+\), there exists another physical eigenstate \(D'_{s1}(2536)^+\), which was first discovered by the ARGUS Collaboration in 1989 \cite{argus}. 
    In the Godfrey-Isgur (GI) relativized constituent quark model \cite{potc, poec}, which does not include explicit tetraquark basis vectors, \(D'_{s1}(2536)^+\) is assigned as the P-wave excited state of the \(c\bar{s}\) system, and its mass and decay behaviors can be well reproduced by the bare \(c\bar{s}\) wave function. 
    In this context, given that experimental data for its radiative decays are still lacking, we also calculate the radiative decay widths of \(D'_{s1}(2536)^+\) in this work to provide a theoretical reference for future experimental measurements.

    Although this research strategy is physically highly motivated, it must be noted that current theoretical methods for calculating radiative decay widths still suffer from inherent limitations. Specifically, within the computational framework, the electromagnetic transition matrix elements typically rely on approximations such as the multipole expansion of plane waves or momentum substitution, while the wave functions are often constructed via a harmonic oscillator basis \cite{ttpp,auat,poec}. Such simplifications inevitably introduce theoretical uncertainties \cite{gemf,rdot}. 
    Therefore, it is necessary to establish a matrix-element calculation scheme without additional simplifying approximations and to introduce a more flexible wave-function expansion method, so as to achieve high-precision quantitative descriptions for radiative decay processes.
	
	In this work, we modify this theoretical method: retaining the complete plane-wave phase factor of the photon to evaluate the radiative transition matrix elements, thereby avoiding simplifying approximations such as multipole expansion and momentum substitution \cite{2790}. Meanwhile, we adopt the Gaussian Expansion Method (GEM) to construct wave functions, which can accurately characterize the asymptotic behaviors in the short- and long-range regions better than harmonic-oscillator bases \cite{gemf}. Based on the above methods, we calculate the mass spectra and radiative decay widths within the pure two-quark picture, and compare our numerical results with the latest experimental data to examine the validity of the two-quark description for these states, which can provide a benchmark for further studies considering two-quark and four-quark mixing. 
	
	The structure of this paper is organized as follows. Following the introduction, Section~\ref{sec:Gaussian} provides an introduction to the ChQM, the GEM, and the framework for calculating electromagnetic decays. Subsequently, Sec.~\ref{sec:discussion} presents the numerical results and related discussions. Finally, Sec.~\ref{sec:Summary} summarizes the main conclusions of this work.
  \section{\label{sec:Gaussian}Quark model and calculation methods}
    In this work, within the framework of the ChQM \cite{qmod,cqms}, we calculate the mass spectra and wave functions of \(D^*_{s0}(2317)^+\) and \(D_{s1}(2460)^+\) using the GEM, and further investigate their radiative decay widths.
	\subsection{The chiral quark  model}
    As a typical phenomenological model for the study of hadron spectra, the ChQM has been employed in numerous works to investigate hadron‑hadron interactions and multiquark systems\cite{qmso,teoh,pdsw,irfa,pops}. The Hamiltonian of this model consists of the mass term, kinetic energy term, center-of-mass kinetic energy term (\(T_c\)), confinement potential (\(V_{\text{CON}}(r_{ij})\)), one-gluon-exchange potential (\(V_{\text{OGE}}(r_{ij})\)), one-\(\sigma\)-meson exchange potential (\(V_{\sigma}(r_{ij})\)), and Goldstone-boson-exchange potential (\(V_{\text{OBE}}(r_{ij})\)). The explicit expressions for these terms are given below:
\begin{equation}
	H=\sum_{i=1}^2\left(m_i+\frac{\boldsymbol{p}_i^2}{2m_i}\right)-T_c+\sum_{i<j}V_{ij}
	\label{eq:Hamiltonian}
\end{equation}
    where
\begin{equation}
	V_{ij} = V_{\text{CON}}(r_{ij})+V_{\text{OGE}}(r_{ij}) + V_{\sigma}(r_{ij})+V_{\text{OBE}}(r_{ij}) \\
\end{equation}
	Here, \(m_i\) and \(\boldsymbol{p}_i\) denote the mass and momentum of the \(i\)-th quark (antiquark), respectively, and \(\boldsymbol{r}_{ij}\) is the relative position between quark \(i\) and quark \(j\). The confinement potential \(V_{\text{CON}}(r_{ij})\) adopts a quadratic form and consists of the central term \(V_\text{CON}^\text{C}(r_{ij})\) and the spin-orbit term \(V_\text{CON}^\text{SO}(r_{ij})\).
\begin{align}
	V_{\text{CON}}^{\text{C}}(r_{ij})
	   &= -a_c \boldsymbol{\lambda}_i^c\cdot\boldsymbol{\lambda}_j^c \big[r_{ij}^2 + \Delta\big]  \\
    V_{\text{CON}}^{\text{SO}}(r_{ij}) 
       &= a_c \boldsymbol{\lambda}_i^c\cdot\boldsymbol{\lambda}_j^c
       \frac{1}{8m_q^2} \frac{1}{r_{ij}} \frac{dV^{\text{c}}}{dr_{ij}} \times \big[\boldsymbol{r}_{ij}  \nonumber\\
       &\quad\times(\boldsymbol{p}_i-\boldsymbol{p}_j)\big]  
       \cdot(\boldsymbol{\sigma}_i+\boldsymbol{\sigma}_j) , ~~~~V^{\text{c}} = r_{ij}^2
	\label{eq:V_CON}
\end{align}
    where \(a_\text{c}\) is the strength parameter of the confinement potential, \(\boldsymbol{\lambda}_i^c\) and \(\boldsymbol{\lambda}_j^c\) are the SU(3) color Gell-Mann matrices, \(\Delta\) is a global constant, and \(\boldsymbol{\sigma}_i\) and \(\boldsymbol{\sigma}_j\) are the spin operators. Unlike the confinement potential, the one-gluon-exchange potential dominates the short-range interaction and consists of the central term \(V_\text{OGE}^\text{C}\), the spin-orbit coupling term \(V_\text{OGE}^\text{SO}\), and the tensor term \(V_\text{OGE}^\text{T}\).
\begin{align}
	V_{\text{OGE}}^{\text{C}}(r_{ij})
    	&= \frac{\alpha_{s}}{4} \boldsymbol{\lambda}_i^c \cdot \boldsymbol{\lambda}_j^c \big[ \frac{1}{r_{ij}}
    	- \frac{\boldsymbol{\sigma}_i \cdot \boldsymbol{\sigma}_j}{6 m_i m_j} 
       \frac{e^{-r_{ij}/r_0(\mu)}}{r_{ij}\,r_0^2(\mu)} \big]     \\
    V_{\text{OGE}}^{\text{SO}}(r_{ij})
        &=-\frac{\alpha_{s}}{16\,m_i^2 m_j^2}\boldsymbol{\lambda}_i^c \cdot \boldsymbol{\lambda}_j^c
        \big[\frac{1}{r_{ij}^3} - \frac{e^{-r_{ij}/r_g(\mu)}}{r_{ij}^3}\big(1    \nonumber \\
        &\quad +\frac{r_{ij}}{r_g(\mu)}\big)\big] \times  \big[\big(m^2_i+m^2_j + 4 m_i m_j \big)(\boldsymbol{S}_+    \nonumber \\
        &\quad \cdot\boldsymbol{L}\big)\big]
        + (m_j^2 - m_i^2)(\boldsymbol{S}_-\cdot\boldsymbol{L})\big]    \\
     V_{\text{OGE}}^{\text{T}}(r_{ij})
        &=-\frac{\alpha_{s}}{16\,m_i m_j}\boldsymbol{\lambda}_i^c \cdot \boldsymbol{\lambda}_j^c
        \big[\frac{1}{r_{ij}^3} - \frac{e^{-r_{ij}/r_g(\mu)}}{r_{ij}^3} \big(\frac{1}{r_{ij}^2}    \nonumber \\
        &\quad + \frac{1}{3\,r_g^2(\mu)}+ \frac{1}{r_{ij}\,r_g(\mu)}\big)\big] S_{ij}   \\
     S_{ij}
        &= 3\frac{(\boldsymbol{\sigma}_i \cdot \boldsymbol{r}_{ij})(\boldsymbol{\sigma}_j \cdot \boldsymbol{r}_{ij})}{r_{ij}^2} - \boldsymbol{\sigma}_i \cdot \boldsymbol{\sigma}_j  \\
     \alpha_s(\mu_{ij}) 
        &= \frac{\alpha_0}{\ln\left[(\mu_{ij}^2 + \mu_0^2)/\Lambda_0^2\right]}
     \label{eq:V_OGE}
\end{align}
    where \(\alpha_{s}\) is the effective strong coupling constant, \(\boldsymbol{S}_\pm = \boldsymbol{S}_i \pm \boldsymbol{S}_j\), \(\boldsymbol{S}_{ij}\) is the tensor force operator, and \(r_g(\mu_{ij}) = \dfrac{r_g}{\mu_{ij}}\). The one-\(\sigma\)-meson exchange provides the intermediate-range attractive interaction and consists of two parts: the central term \(V_\sigma^\text{C}(r_{ij})\) and the spin-orbit coupling term \(V_\sigma^\text{SO}(r_{ij})\).
\begin{align}
	V_\sigma^\text{C}(r_{ij})
		&=-\frac{g^{2}_{\text{ch}}}{4\pi}\frac{\Lambda_{\sigma}^{2}m_{\sigma}}{\Lambda_{\sigma}^{2}-m_{\sigma}^{2}}
		\big[Y(m_\sigma r_{ij}) - \frac{\Lambda_{\sigma}}{m_{\sigma}}Y(\Lambda_{\sigma} r_{ij})\big]   \\
   	V_\sigma^\text{SO}(r_{ij})
   	    &=-\frac{g^{2}_{\text{ch}}}{4\pi}\frac{\Lambda_{\sigma}^2}{\Lambda_{\sigma}^2 - m_{\sigma}^2}\frac{m_{\sigma}^3}{2m_i m_j}\big[G(m_{\sigma} r_{ij})   \nonumber\\
   	    &\quad - \big(\frac{\Lambda_{\sigma}}{m_{\sigma}}\big)^3 G(\Lambda_{\sigma} r_{ij})
   	    \big]\boldsymbol{L}\cdot\boldsymbol{S}  
   	\label{eq:V_sigma}
\end{align}
    where the standard Yukawa function \(Y(x)\) and its related functions are defined as \(Y(x) = e^{-x}/x\), \(G(x) = (1+1/x)Y(x)/x\), and \(H(x) = (1+3/x+3/x^2)Y(x)/x\) \cite{cqms}.Goldstone-boson exchange describes the long-range chiral interaction, including the central term \(V_{\chi=\pi,\text{K},\eta}^\text{C}\) and the tensor term \(V_{\chi=\pi,\text{K},\eta}^\text{T}\).
\begin{align}
    V^{\text{OBE}}(r_{ij}) 
       &=V_{\pi}(r_{ij})\sum_{a=1}^3 \boldsymbol{\lambda}_i^a \cdot \boldsymbol{\lambda}_j^a
       + V_{K}(r_{ij})\sum_{a=4}^7 \boldsymbol{\lambda}_i^a \cdot \boldsymbol{\lambda}_j^a    \nonumber \\
       &\quad + V_{\eta}(r_{ij})\left[\left(\boldsymbol{\lambda}_i^8 \cdot \boldsymbol{\lambda}_j^8\right)\cos\theta_p - \left(\boldsymbol{\lambda}_i^0 \cdot \boldsymbol{\lambda}_j^0\right)\sin\theta_p \right]     \\
   V_{\chi=\pi,\text{K},\eta}^\text{C}(r_{ij}) 
       &= \frac{g^{2}_{\text{ch}}}{4\pi}\frac{m_{\chi}^2}{12m_{i}m_{j}}
       \frac{\Lambda_{\chi}^2 m_\chi}{\Lambda_{\chi}^2-m_{\chi}^2} \big[Y(m_{\chi} r_{ij}) -   \frac{\Lambda_{\chi}^3}{m_{\chi}^3}
       \nonumber\\ 
       &\quad Y(\Lambda_{\chi} r_{ij})\big]  \left(\boldsymbol{\sigma}_i \cdot \boldsymbol{\sigma}_j\right)      \\
   V_{\chi=\pi,\text{K},\eta}^\text{T}(r_{ij}) 
       &= \frac{g^{2}_{\text{ch}}}{4\pi}\frac{m_{\chi}^2}{12m_{i}m_{j}}
       \frac{\Lambda_{\chi}^2 m_\chi}{\Lambda_{\chi}^2-m_{\chi}^2}\big[H(m_{\chi} r_{ij}) -   \frac{\Lambda_{\chi}^3}{m_{\chi}^3}
       \nonumber\\ 
       &\quad H(\Lambda_{\chi} r_{ij})\big] S_{ij}  
   \label{eq:V_OBE}
\end{align}
    where \(\boldsymbol{\lambda}_i^a\) and \(\boldsymbol{\lambda}_j^a\) are the Gell-Mann matrices in SU(3) flavor space, \(\Lambda_{\chi(\sigma)}\) is the cutoff parameter, \(g_\text{ch}\) is the chiral coupling constant, and \(\theta_P\) denotes the mixing angle between \(\eta_0\) and \(\eta_8\). These parameters are determined by fitting the ground-state meson spectra. The specific parameter values are listed in Table~\ref{quarkpara}, and the corresponding fitted meson spectra are presented in Table~\ref{masses}. 
\begin{table}[htb]
	\caption{Quark model parameters ($m_\pi = 0.7\ \mathrm{fm}^{-1}$, $m_\sigma = 3.42\ \mathrm{fm}^{-1}$, $m_\eta = 2.77\ \mathrm{fm}^{-1}$,$m_K = 2.51\ \mathrm{fm}^{-1}$)}
	\label{quarkpara}
	\begin{ruledtabular}
		\begin{tabular}{lcc}
			Quark masses & $m_u=m_d$(MeV) & 313 \\
			& $m_s$(MeV) & 536 \\
			& $m_c$(MeV) & 1728 \\
			\hline
			Goldstone bosons & $\Lambda_\pi$=$\Lambda_\sigma(\mathrm{fm}^{-1})$ & 4.2 \\
			& $\Lambda_\eta$=$\Lambda_K(\mathrm{fm}^{-1})$ & 5.2 \\
			& $g_\mathrm{ch}^2/(4\pi)$ & 0.54 \\
			& $\theta_p(^\circ)$ & -15 \\
			\hline
			Confinement & $a_c(\mathrm{MeV\cdot fm^{-2}})$ & 101 \\
			& $\Delta(\mathrm{MeV})$ & -78.3 \\
			\hline
			OGE & $\Lambda_{0}(\mathrm{fm}^{-1})$ & 0.033 \\
			& $\mu_{0}(\mathrm{MeV})$ & 36.976 \\
			& $\hat{r}_0(\mathrm{MeV})$ & 28.17 \\
			& $\hat{r}_g(\mathrm{MeV})$ & 34.5 \\
			& $\alpha_{0}$ & 3.67 \\
			& $a_s$ & 0.777 \\
		\end{tabular}
	\end{ruledtabular}
\end{table}

\begin{table}[H]
	\centering
	\caption{The calculated masses (in MeV) of the mesons in
		ChQM.Experimental values are taken from the Particle Data
		Group (PDG) \cite{PDG}}
	\label{masses}
	\begin{tabular}{c*{10}{c}}  
		\hline
		\hline
		& $\pi$ & $\rho$ &$K$ &$K^*$ &$\phi$ &$\omega$ & $D$ & $D^*$ & $D_s$ & $D^*_s$   \\
		\hline
		\text{ChQM} & 139 & 773 & 494 & 914 & 1016 & 702 & 1863 & 1981 & 1953 & 2080  \\
		\text{Exp.} & 140 & 775 & 494 & 892 & 1020 & 783 & 1865 & 2007 & 1968 & 2112  \\
		\hline
	\end{tabular}
\end{table}

	\subsection{Gaussian expansion method}
	GEM was originally proposed by Kamimura within a non-adiabatic framework to solve three-body problems of \(\mu\)-atoms and \(\mu\)-molecules \cite{crca}. Subsequently, Hiyama, Kino, and Kamimura systematically extended this method and established a complete theoretical framework for GEM \cite{gemf}. This method employs Gaussian basis functions distributed in a geometric progression as the variational basis set to construct orbital wave functions, thereby rapidly converging the eigenenergies of the stationary Schrödinger equation for few-body systems to a specified accuracy. It has been widely applied to mass spectrum calculations for mesons, baryons, and multiquark states \cite{tcab,dsoa,esit,eacc,btia,bane,dsot}. In this work, for the meson systems, we expand the orbital wave function into a set of Gaussian wave functions distributed in geometric progression, namely:
\begin{align}
    \psi_{lm}(\boldsymbol{r}) 
       &= \sum_{n=1}^{n_{\text{max}}} c_{nl} \phi_{nlm}(\boldsymbol{r}) \\
    \phi_{nlm}(\boldsymbol{r}) 
       &= \phi_{nl}(r)\, Y_{lm}(\hat{\boldsymbol{r}})  \\
    \phi_{nl}(r) 
       &= N_{nl}\, r^{l} e^{-\nu_{n} r^{2}}  
	\label{eq:GEM}
\end{align}
    In this equation, \(c_{nl}\) are the variational parameters, which denote the expansion coefficients of the basis functions \(\phi_{nlm}(\boldsymbol{r})\), \(N_{nl}\) are the normalization constants,
\begin{equation}
    N_{nl} = \big( \frac{2^{l+2}\,(2\nu_{n})^{l+3/2}}{\sqrt{\pi}\,(2l+1)!!} \big)^{\frac{1}{2}}
    \label{eq:N_nl}
\end{equation}
    $n$ runs from $1$ to $n_{\mathrm{max}}$, where $n_{\mathrm{max}}$ denotes the total number of basis functions adopted in the Gaussian expansion, $a$ is the geometric scaling factor satisfying:
\begin{equation}
    \nu_n = \frac{1}{r_n^2} , 
    r_n = r_{\mathrm{min}} a^{n-1} , 
    a = \big(\frac{r_{n_{\mathrm{max}}}}{r_{n_{\mathrm{min}}}}\big)^{\frac{1}{n_{\mathrm{max}}-1}}\\
    \label{eq:nu_n}
\end{equation}
    To ensure the convergence of the numerical results, we adopt 8 Gaussian basis functions, with Gaussian widths ranging from \(0.1\ \mathrm{fm}\) to \(2\ \mathrm{fm}\). Within the combined framework of ChQM and GEM, the obtained meson mass spectra and wave functions serve as the essential input for the subsequent calculation of radiative decay widths.

    \subsection{Electromagnetic decay}
    The quark-photon coupling Hamiltonian is given by
\begin{equation}
	H_\mathrm{int}=-\sum_j e_j \bar{\psi}_j \gamma^\mu A_\mu \psi_j
	\label{eq:H_int}
\end{equation}
    where \(e_j\) and \(\psi_j\) denote the electric charge and quark field of the \(j\)-th quark (antiquark), respectively. The Dirac conjugate of \(\psi_j\) is defined as \(\bar{\psi}_j = \psi_j^\dagger \gamma^0\), with \(\gamma^0=\begin{pmatrix}1 & 0 \\ 0 & -1\end{pmatrix}\) and \(\gamma^i=\begin{pmatrix}0 & \sigma^i \\ -\sigma^i & 0\end{pmatrix}\). Here, \(\gamma^\mu\) are \(4\times4\) matrices and \(\sigma^i\) are the \(2\times2\) Pauli matrices. In the nonrelativistic limit, the Hamiltonian takes the following form:
\begin{align}
    H_{\text{em}} 
       &= \sum_{j=1}^{2} \frac{e_j}{\sqrt{2k}}
       \big[-\varepsilon_0 e^{-i\boldsymbol{k}\cdot\boldsymbol{r}_j}
       - \frac{\boldsymbol{p}_j\cdot\boldsymbol{\varepsilon}\, e^{-i\boldsymbol{k}\cdot\boldsymbol{r}_j} + e^{-i\boldsymbol{k}\cdot\boldsymbol{r}_j}\boldsymbol{\varepsilon}\cdot\boldsymbol{p}_j}{2m_j}
       \nonumber\\
       &\quad -\frac{i\boldsymbol{\sigma}\cdot(\boldsymbol{k}\times\boldsymbol{\varepsilon})\,e^{-i\boldsymbol{k}\cdot\boldsymbol{r}_j}}{2m_j}
       \big]
       \label{eq:H_em}
\end{align}
    where the summation index \(j\) labels the constituent quarks inside the hadron, and \(m_j\) is the constituent mass of the j-th quark, and \(\boldsymbol{r}_j\) and \(\boldsymbol{p}_j\) correspond to its spatial coordinate and canonical momentum. The photon propagates along the \(+z\) direction with momentum \(\boldsymbol{k}=k\hat{\boldsymbol{z}}\), and its magnitude \(k\) equals the photon energy, \(e^{-i\boldsymbol{k}\cdot\boldsymbol{r}_j}\) is the plane-wave phase factor of the photon field evaluated at the quark position \(\boldsymbol{r}_j\), and \(\varepsilon_\mu=(0,1,-i,0)/\sqrt{2}\) is the photon polarization four-vector. Since the radiative decay process involves only real photons, the time component of the polarization vector satisfies \(\varepsilon_0=0\), which makes the first term of the Hamiltonian vanish identically. The second term corresponds to the orbital electromagnetic coupling, while the third term corresponds to the spin-photon coupling. Substituting the above photon kinematic conditions and simplifying, the Hamiltonian finally reduces to:
\begin{align}
	H_{\text{em}} 
	   &= \sum_{j=1}^{2} \sqrt{\frac{4\pi\alpha}{k}} \mu_j
	   \Big[ k\boldsymbol{s}_{j,-}\, e^{-i\boldsymbol{k}\cdot\boldsymbol{r}_j} \nonumber \\
	   &\quad - \frac12 \big(\boldsymbol{p}_{j,-} e^{-i\boldsymbol{k}\cdot\boldsymbol{r}_j}
	   + e^{-i\boldsymbol{k}\cdot\boldsymbol{r}_j}\boldsymbol{p}_{j,-}\big) \Big]
	\label{eq:H_em2}
\end{align}
    where \(\mu_j=q_j/(2m_j)\), and \(q_j\) is the dimensionless charge fraction satisfying \(q_j=e_j/e\), with the elementary charge defined as \(e=\sqrt{4\pi\alpha}\) in natural units. \(\boldsymbol{s}_{j,-}=s_{j,x}-is_{j,y}\) and \(\boldsymbol{p}_{j,-}=p_{j,x}-ip_{j,y}\) are the spin-lowering and momentum-lowering operators of the \(j\)-th quark, respectively. For notational simplicity, we introduce the shorthand operators \(\hat{U}_j \equiv e^{-i\boldsymbol{k}\cdot\boldsymbol{r}_j}\) and \(\hat{\boldsymbol{T}}_{j,-}\equiv \boldsymbol{p}_{j,-}e^{-i\boldsymbol{k}\cdot\boldsymbol{r}_j}+e^{-i\boldsymbol{k}\cdot\boldsymbol{r}_j}\boldsymbol{p}_{j,-}\) \cite{sdwa}. Using the above definitions, the Hamiltonian can be rewritten as \cite{rdot}:
\begin{align}
	H_{\text{em}}
	   &= \sum_{j=1}^{2} 2\sqrt{\frac{\pi\alpha}{k}} \mu_j
	   \big[k \boldsymbol{s}_{j,-}\, \hat{U}_j - \frac12 \hat{\boldsymbol{T}}_{j,-}\big]
	\label{eq:H_em3}
\end{align}
    By evaluating the transition matrix element of the electromagnetic Hamiltonian \(H_{\mathrm{em}}\) between the initial and final meson states, the helicity amplitude \(A_{M_{J_A}}\) is obtained as:
\begin{align}
	A_{M_{J_A}}
	   &= \langle \phi_{A'},\, k_{A'},\, J_{A'},\, M_{J_{A'}} \vert H_{\text{em}} \vert \phi_A,\, k_A,\, J_A,\, M_{J_A} \rangle
	\label{eq:A_MJA}
\end{align}
    where \(\vert \phi_A,\, k_A,\, J_A,\, M_{J_A} \rangle\) and \(\vert \phi_{A'},\, k_{A'},\, J_{A'},\, M_{J_{A'}} \rangle\) denote the initial and final states, respectively. \(\phi_A\) represents the flavor wave function, \(k_A\) labels the basis function node, \(J_A\) is the total angular momentum of the system, and \(M_{J_A}\) is the projection of the total angular momentum along the \(z\)-axis. The photon emission process imposes the magnetic quantum number selection rule: \(M_{J_{A'}} = M_{J_A}-1\). The detailed derivation and treatment of the helicity amplitude \(A_{M_{J_A}}\) are provided in Appendix~\ref{Helicity Amplitude}.
    
    Following Fermi's golden rule, the radiative decay width is obtained by substituting the helicity amplitude Eq.~\eqref{eq:A_MJA} into Eq.~\eqref{eq:Gamma}, which is given below:
\begin{align}
	\Gamma 
	   &= \frac{k^2}{\pi}\cdot \frac{2}{2J_A+1}\cdot \frac{E_A'}{m_A}\cdot \sum_{M_{J_A}>0} |A_{M_{J_A}}|^2
	\label{eq:Gamma}
\end{align}
    where the photon energy is fixed by energy-momentum conservation for the two-body decay as \(k=(m_A^2-m_{A'}^2)/(2m_A)\), \(E_{A'}\) is the relativistic energy of the final meson satisfying \(E_{A'}=\sqrt{m_{A'}^2+k^2}\), and \(m_A\), \(m_{A'}\) denote the masses of the initial and final mesons, respectively.
    
    It is worth noting that, in deriving the radiative transition amplitudes, we do not perform a multipole expansion of the plane-wave factor. Instead, we retain the complete exponential phase factor and carry out the integration directly. Furthermore, we do not adopt the approximation commonly used in some literature, namely substituting \(\boldsymbol p_j/m_j\) with \(-i\boldsymbol k\cdot\boldsymbol r_j\) via the commutation relation \(i[H_0,\boldsymbol r_j]=\boldsymbol p_j/m_j\) \cite{close}. By avoiding this approximation, we obtain exact analytical transition amplitudes.

  \section{\label{sec:discussion}The results and discussions}
    In this work, we treat the three charmed-strange states \(D^*_{s0}(2317)^+\), \(D_{s1}(2460)^+\), and \(D'_{s1}(2536)^+\) as conventional \(c\bar{s}\) structures. By applying the ChQM combined with the GEM, we first calculate their mass spectra. Using the obtained wave functions, we then evaluate their radiative decay widths.
    Based on these observables, we compare our theoretical results with the experimental measurements and other theoretical results, and further discuss the possible internal configurations of these $D_{sJ}$ states.
    The results are listed in Table~\ref{2317}.
    
     For the \(0^+\) state, the calculated mass is 2337~MeV, which is relatively close to the experimental \(D^*_{s0}(2317)^+\) and can be used to interpret this state. For the \(1^+\) states, the pure \(1^1P_1\) and \(1^3P_1\) configurations yield masses of 2456~MeV and 2479~MeV, respectively. The former is close to \(D_{s1}(2460)^+\), while the latter is 57~MeV lower than \(D'_{s1}(2536)^+\). In fact, the \(1^1P_1\) and \(1^3P_1\) states can mix with each other. The mixing scheme adopted in this work is as follows \cite{maba}:
    \begin{align}
    	P &= {}^1P_1 \cos\theta_{nP} + {}^3P_1 \sin\theta_{nP}   \nonumber \\
    	P' &= -^1P_1 \sin\theta_{nP} + {}^3P_1 \cos\theta_{nP}
    \end{align}
    The mixing angle obtained self-consistently within the model is \(\theta_{nP}=56.17^\circ\). After introducing the mixing, the masses of the resulting two physical eigenstates are 2434~MeV and 2497~MeV, which can simultaneously describe \(D_{s1}(2460)^+\) and \(D'_{s1}(2536)^+\), respectively.
    \setlength{\tabcolsep}{3pt}
\begin{table}[H]
	\centering
	\caption{The masses (in MeV) of the \(D^*_{s0}(2317)^+\), \(D_{s1}(2460)^+\), and \(D'_{s1}(2536)^+\) mesons in ChQM. For comparison, we also list the theoretical estimations in Ref.~\cite{etot,tpop,qcpm} and experimental measurements \cite{PDG}.}
	\label{2317}
	\begin{tabular}{c*{5}{c}}  
		\hline
		\hline
		& Ours & Ref.~\cite{etot} & Ref.~\cite{tpop} & Ref.~\cite{qcpm} & Exp.   \\
		\hline
		\(D^*_{s0}(2317)^+\)  & 2337 & 2317 & 2480 & 2388 & 2317.8   \\
		\(D_{s1}(2460)^+\)  & 2434 & 2425 & 2550 & 2521 & 2459.5   \\
		\(D'_{s1}(2536)^+\) & 2497 & 2510 & 2560 & 2537 & 2535.1   \\
		\hline
	\end{tabular}
\end{table}
    Here, we list several theoretical works that also treat the $D_{sJ}$ states as $c\bar{s}$ configurations in the table. It is obvious that our work is close to Ref.~\cite{etot}, which adopts the relativistic constituent quark model including spin-spin and spin-orbit coupling interactions, and incorporates the $S$-$D$ mixing as well as singlet-triplet spin mixing effects. Ref.~\cite{tpop} uses a relativized quark model to calculate the mass spectra of singlet P-wave mesons, constructs an effective potential with momentum-coordinate coupling, and implements the relativistic correction of the potential via Gaussian broadening and substitution of the inverse mass factor. 
    While Ref.~\cite{qcpm} employs the Quantum‑Chromodynamics potential model to investigate the mass spectra of light-heavy quark systems, incorporates one-loop radiative corrections into the quark-antiquark potential and includes spin-orbit mixing terms. 
    Notably, Ref.~\cite{sotn} treats \(D^*_{s0}(2317)^+\) and \(D_{s1}(2460)^+\) as compact tetraquark states. Although their initial mass spectra calculated with universal parameters are about 150~MeV higher than the experimental values, the masses shift to 2317.87~MeV and 2459.58~MeV after parameter refitting, which agree perfectly with the data. This indicates that it is difficult to determine the internal structures of \(D_{sJ}\) states based solely on mass spectra. Therefore, it is necessary to further investigate their radiative decays.
    
    Then, using the meson wave functions constrained by the mass spectra, we further calculate the radiative decay widths of these resonances, with the theoretical mass values taken from Table~\ref{2317}.
    The calculated results from other references and available experimental constraints are summarized in Table~\ref{width}. Note that the results from Ref.~\cite{potc,pmca,poec,cmoh,io2317} are obtained within the conventional constituent-quark-model \(c\bar{s}\) meson picture, whereas those from Ref.~\cite{uosa,sard} are calculated within the tetraquark theoretical framework.
    \begin{table*}[t!]
    	\caption{\raggedright Radiative partial widths of $D^*_{s0}(2317)^+$, $D_{s1}(2460)^+$, and $D'_{s1}(2536)^+$. The Exp. column summarizes the available experimental constraints, which are taken from the two experimental papers in Refs.~\cite{draft,ootr}.}
    	\label{width}
    	\begin{tabular}{l *{10}{c}}
    		\hline\hline
    		Decay channel &Ours  & Ref.~\cite{potc} & Ref.~\cite{pmca} & Ref.~\cite{poec} & Ref.~\cite{cmoh} & Ref.~\cite{io2317} & Ref.~\cite{uosa} & Ref.~\cite{sard} & Exp. &  \\
    		\hline
    		$\Gamma(D^*_{s0}(2337)^+ \to D_s^{*}\gamma)$ [keV] &3.80 &1.90 &4.92 &9.01 &1.74 &1.00 &3.70 &9.40  &$\cdots$ & \\
    		$\Gamma(D^*_{s0}(2337)^+ \to D_s\pi)$ [keV] &$\cdots$ &10.00 &$\cdots$ &$\cdots$ &21.50 &7.00 &132.00 &96.00 &$\cdots$ & \\
    		Ratio=$\dfrac{\Gamma(D^*_{s0}(2337)^+ \to D_s^{*}\gamma)}{\Gamma(D^*_{s0}(2337)^+ \to D_s\pi)}$ &$\cdots$ &0.19 &$\cdots$ &$\cdots$ &0.0809 &0.1429 &0.028 &0.0979 &0.0714 & \\
    		\hline
    		$\Gamma(D_{s1}(2434)^+ \to D_s^{*}\gamma)$ [keV] &4.97 &5.50 &15.50 &5.40 &4.66 &1.50 &13.00 &25.20 &$\cdots$ & \\
    		$\Gamma(D_{s1}(2434)^+ \to D_s^\gamma)$ [keV] &8.65 &6.20 &12.80 &15.20 &5.08 &3.30 &42.00 &24.20 &$\cdots$ & \\
    		Ratio=$\dfrac{\Gamma(D_{s1}(2434)^+ \to D_s^{*}\gamma)}{\Gamma(D_{s1}(2434)^+ \to D_s\gamma)}$ &0.57 &0.89 &1.21 &0.36 &0.92 &0.45 &0.31 &1.04 &0.74 & \\
    		\hline
    		$\Gamma(D'_{s1}(2497)^+ \to D_s^{*}\gamma)$ [keV] &9.14 &5.60 &8.90 &9.61 &$\cdots$ &$\cdots$ &$\cdots$ &$\cdots$ &$\cdots$ & \\
    		$\Gamma(D'_{s1}(2497)^+ \to D_s^{}\gamma)$ [keV] &8.59 &15.00 &54.5 &9.23 &$\cdots$ &$\cdots$ &$\cdots$ &$\cdots$ &$\cdots$ & \\
    		Ratio=$\dfrac{\Gamma(D'_{s1}(2497)^+ \to D_s^{*}\gamma)}{\Gamma(D'_{s1}(2497)^+ \to D_s\gamma)}$ &1.06 &0.37  &0.16 &1.04 &$\cdots$ &$\cdots$ &$\cdots$ &$\cdots$ &$\cdots$ &  \\
    		\hline
    		$\Gamma(D_{s1}(2456)^+ \to D_s^{*}\gamma)$ [keV] &4.16 & & & & & & & & & \\
    		$\Gamma(D_{s1}(2456)^+ \to D_s^\gamma)$ [keV] &9.54 & & & & & & & & & \\
    		Ratio=$\dfrac{\Gamma(D_{s1}(2456)^+ \to D_s^{*}\gamma)}{\Gamma(D_{s1}(2456)^+ \to D_s\gamma)}$ &0.44 & & & & & & & & & \\
    		\hline
    		$\Gamma(D_{s1}(2479)^+ \to D_s^{*}\gamma)$ [keV] &10.73 & & & & & & & & & \\
    		$\Gamma(D_{s1}(2479)^+ \to D_s^\gamma)$ [keV] &7.89 & & & & & & & & & \\
    		Ratio=$\dfrac{\Gamma((D_{s1}(2479)^+ \to D_s^{*}\gamma)}{\Gamma((D_{s1}(2479)^+ \to D_s\gamma)}$ &1.36 & & & & & & & & &  \\
    		\hline\hline
    	\end{tabular}
    \end{table*}
    
    As shown in Table~\ref{width}, the width of the radiative decay \(D^*_{s0}(2317)^+ \to D_s^{*+}\gamma\) is 3.80~keV. This result is consistent with other theoretical calculations, which fall in the range of 1.0--9.4~keV, all on the order of several keV. 
    For the strong decay of \(D^*_{s0}(2317)^+\), its exact value is difficult to determine due to isospin breaking, leading to significant discrepancies among theoretical works, with calculated results ranging from 7.0 to 132~keV. Taking the ratio of our calculated radiative width to the theoretical strong decay widths from other works, we obtain a value of 0.029--0.54, whereas the experimentally measured branching-fraction ratio is 0.0714. 
    Furthermore, following the experimental work in Ref.~\cite{ootr}, Ref.~\cite{tdta} treats \(D_{s0}^*(2317)^+\) as a molecular state and reports the decay widths of \(D_{s0}^*(2317)^+ \to D_s^*\gamma\) and \(D_{s0}^*(2317)^+ \to D_s\pi\) as 1.33~keV and 70.8~keV, respectively. Taking the ratio of our calculated radiative width to the strong decay width from Ref.~\cite{tdta}, we obtain a value of 0.053, which is closer to the experimental measurement.
    
    For \(D_{s1}(2460)^+\), we treat it as a mixed state of \(1^1P_1\) and \(1^3P_1\). For the theoretical mass of 2434~MeV, the radiative decay widths to \(D_s^{*+}\gamma\) and \(D_s^+\gamma\) are calculated to be 4.97~keV and 8.65~keV, respectively, yielding a ratio of 0.57. Compared with other theoretical works, we find that their ratios fall in the range of 0.31--1.21, while the experimental value is 0.74. The possible factors for these discrepancies include the theoretical methods and the choice of mixing angles (our self-consistently determined mixing angle is \(56.17^\circ\)). 
    For instance, Ref.~\cite{pmca} obtained a mixing angle of \(59.7^\circ\) using a semi-relativistic potential model, whereas Ref.~\cite{poec} obtained \(-37.48^\circ\) using the GI model, leading to inconsistent results for the radiative widths. Furthermore, although both Refs.~\cite{potc} and~\cite{poec} employ the GI model, the former directly inputs the experimental masses (2317 and 2460~MeV), while the latter uses theoretical parameters, resulting in divergent outcomes. Meanwhile, Refs.~\cite{uosa} and~\cite{sard} treat them as molecular states, but their results also show considerable deviations from the experimental data.
    
    For comparison, we also calculate the width ratios for the pure \(1^3P_1\) and pure \(1^1P_1\) states, which are 0.44 and 1.36, respectively, and both of them deviate even more from the experiment. 
    Therefore, we prefer to interpret \(D_{s1}(2460)^+\) as a mixed state of \(1^1P_1\) and \(1^3P_1\). 
    Certainly, the above calculations are based on the conventional \(c\bar{s}\) structure. 
    In fact, \(D_{s1}(2460)^+\), \(D_s^*\), and \(D_s\) may all contain tetraquark components. Calculations in Ref.~\cite{ssod} based on ChQM show that the tetraquark components in these three states are 42.4\%, 2.3\%, and 1.5\%, respectively. 
    On the one hand, the tetraquark component in \(D_s^{*}\) exceeds that in \(D_s\); on the other hand, the decay width of \(D_{s1}(2460)^+\) is generally larger in the tetraquark picture than in the conventional meson picture. 
    Therefore, incorporating tetraquark components for \(D_{s1}(2460)^+\) would yield a ratio greater than 0.57.
    Based on the above analysis, we tend to conclude that \(D_{s1}(2460)^+\) is a mixed state containing tetraquark components.

    Furthermore, we obtain a state with a mass of approximately 2497 MeV and a radiative branching ratio of 1.06. We identify this state as \(D'_{s1}(2536)^+\). 
    From Table~\ref{width}, our calculated partial width is \(\Gamma(D'_{s1}(2536)^+\to D_s^{*+}\gamma)=9.14\,\mathrm{keV}\). The results reported in Refs.~\cite{potc,pmca,poec} cover the range 5.6\(-\)9.61~keV. For \(\Gamma(D'_{s1}(2536)^+\to D_s^+\gamma)=8.59\,\mathrm{keV}\), the values given in Refs.~\cite{potc,pmca,poec} fall within the interval 9.23\(-\)54.5~keV. Our results are slightly smaller than those given in the existing literature.
    Ref.~\cite{wdwn} analyzes the published experimental data from CLEO \cite{pado} and BaBar \cite{asot} and obtains \(\mathrm{Br}[D'_{s1}(2536)^+\to D_s^+\gamma]<0.009\). Adopting the total width \(\Gamma_{\mathrm{tot}}=0.92\pm0.05\ \mathrm{MeV}\) of \(D'_{s1}(2536)^+\) from the PDG and using the relation \(\Gamma_{\mathrm{rad}}=\mathrm{Br}\cdot\Gamma_{\mathrm{tot}}\), one can derive \(\Gamma(D'_{s1}(2536)^+\to D_s^+\gamma)<8\ \mathrm{keV}\). It should be emphasized that this value is a phenomenological upper limit extracted from indirect extrapolation, and the true value may fluctuate within a narrow range. Our theoretical calculation \(\Gamma(D'_{s1}(2536)^+\to D_s^+\gamma)=8.59\ \mathrm{keV}\) lies within this fluctuation range. The radiative decay width of the \(D'_{s1}(2536)^+\) state can be well reproduced within a pure \(c\bar{s}\) model. 
    Meanwhile, Ref.~\cite{nccf} adopts an effective Hamiltonian coupled-channel model to fit the finite-volume energy spectra from lattice QCD.
    Their conclusion shows that \(D'_{s1}(2536)^+\) contains a \(c\bar{s}\) component of 98.2\% and is dominated by the bare \(c\bar{s}\) core. These findings suggest that \(D'_{s1}(2536)^+\) could be a mixture of the pure \(c\bar{s}\) meson component and the molecular component. The fraction of the molecular component is rather small and its contribution can be safely neglected.
    Nevertheless, the definitive determination of the internal structure of \(D'_{s1}(2536)^+\) still awaits more precise experimental measurements.

  \section{\label{sec:Summary}Summary}
    In this work, we employ the GEM within the ChQM to calculate the energy spectra and obtain the corresponding wave functions of \(D_{s0}^{*}(2317)^+\), \(D_{s1}(2460)^+\) and \(D'_{s1}(2536)^+\). The radiative decay study is carried out based on these spectra and wave functions. The obtained masses are 2337 MeV, 2434 MeV and 2497 MeV, respectively. The radiative decay widths are
\begin{align*}
	\Gamma(D_{s0}^{*}(2317)^+\to D_s^{*+}\gamma) &= 3.80\ \text{keV} \\
	\Gamma(D_{s1}(2460)^+\to D_s^{*+}\gamma) &= 4.97\ \text{keV} \\
	\Gamma(D_{s1}(2460)^+\to D_s^+\gamma) &= 8.65\ \text{keV} \\
	\Gamma(D'_{s1}(2536)^+\to D_s^{*+}\gamma) &= 9.14\ \text{keV} \\
	\Gamma(D'_{s1}(2536)^+\to D_s^+\gamma) &= 8.59\ \text{keV}
\end{align*}  
    This shows the radiative decay‑width ratios \(\Gamma(D_{s1}(2460)^+\to D_s^{*+}\gamma)/\Gamma(D_{s1}(2460)^+\to D_s^+\gamma)=0.57\) and \(\Gamma(D'_{s1}(2536)^+\to D_s^{*+}\gamma)/\Gamma(D'_{s1}(2536)^+\to D_s^+\gamma)=1.06\).
    Calculations within the pure \(c\bar{s}\) picture indicate that \(D_{s0}^{*}(2317)^+\) and \(D_{s1}(2460)^+\) contain tetraquark components. Spectral calculations for mixed states show that the \(c\bar{s}\) component dominates, and radiative decay analysis further demonstrates that the tetraquark component cannot be neglected \cite{ssod}. 
    Therefore, we tend to interpret these two states as meson–tetraquark mixed states. 
    Some existing theoretical analyses show that the tetraquark fraction of \(D_{s1}'(2536)^+\) is negligible compared with its \(c\bar{s}\) component, so it can be approximated as a pure \(c\bar{s}\) state. 
    However, further experimental tests for the internal structure of \(D_{sJ}\) and precise measurements of radiative‑decay observables are still necessary. 
    To achieve more precise results for \(D_{s0}^{*}(2317)^+\) and \(D_{s1}(2460)^+\), self-consistent simultaneous computations of the mass spectrum and radiative decay within the meson–tetraquark mixed configuration are required.
    
    A large number of charm‑hadron states have been identified, and considerable difficulties persist in probing their internal structures. Different theoretical efforts adopt separate approaches based on mass‑spectrum calculations or hadronic‑decay analyses. To deliver adequate theoretical references for experimental studies, combined theoretical treatments of mass spectra and decay observables are highly necessary. Moreover, the results obtained in the present work cannot satisfactorily reproduce the experimental measurements for \(D_{s0}^{*}(2317)^+\) and \(D_{s1}(2460)^+\). This limitation originates from the fact that our computations are constrained to one single structural configuration. This observation further illustrates that the internal hadronic configuration exerts a decisive impact on decay properties, representing an important research avenue for coordinated theoretical and experimental investigations.
    
	\label{22}
	\appendix
  \section{\label{Helicity Amplitude}Helicity Amplitudes and Momentum Ladder Operators}
	Substituting the electromagnetic transition Hamiltonian given in Eq.~\eqref{eq:H_em3} into the definition of helicity amplitude Eq.~\eqref{eq:A_MJA}, we obtain the transition matrix element for meson radiative decay. 
\begin{align}
	A_{M_{J_A}} 
	   &=  2\sqrt{\frac{\pi\alpha}{k}} \,k \sum_{j=1}^{2} \big[\langle \phi_{A'}, k_{A'}, J_{A'}, M_{J_{A'}} \vert     \nonumber\\
	   &\quad \times \mu_j \boldsymbol{s}_{j,-} \hat{U}_j \vert \phi_A, k_A, J_A, M_{J_A} \rangle\big] \nonumber\\
	   &\quad - \sqrt{\frac{\pi\alpha}{k}} \sum_{j=1}^{2}\big[\langle \phi_{A'}, k_{A'}, J_{A'}, M_{J_{A'}} \vert    \nonumber\\
	   &\quad \times \mu_j \hat{\boldsymbol{T}}_{j,-} \vert \phi_A, k_A, J_A, M_{J_A} \rangle\big]
\end{align}
	This matrix element can be decomposed at the quark level into independent flavor, spin and spatial components. Accordingly, we perform the angular-momentum coupling decomposition via Clebsch-Gordan coefficients and carry out the basis expansion for the initial and final meson wave functions.
\begin{align}
	|\phi_A,k_A,J_A,M_{J_A}\rangle
	   &= \sum_{M_L,M_S} \langle L,M_L;S,M_S\vert J_A,M_{J_A}\rangle \nonumber\\
	   &\times \sum_{m_{l}} \langle l,m_{l}\vert L,M_L\rangle   \nonumber\\
	   &\times |S_1,m_{S_1}\rangle \otimes |S_2,m_{S_2}\rangle    \nonumber\\
 	   &\otimes |\phi_A\rangle \otimes |k,l,m_l\rangle
\end{align}
    where \(|S_1,m_{S_1}\rangle\) and \(|S_2,m_{S_2}\rangle\) are the spin wave functions of the quark and antiquark, respectively, \(|k,l,m_l\rangle\) denotes the spatial wave function of the meson obtained via the GEM, where k is the radial quantum number, l is the orbital angular momentum, and \(m_l\) is its third component.
    
    The matrix element of the tensor operator \(\hat{\boldsymbol{T}}_{j,-}\) is expressed in terms of \(\hat U_j\) and the ladder operator \(\boldsymbol{p}_{j,-}\). 
    To carry out this evaluation, we present the explicit expression of the matrix element in the basis introduced above:
\begin{align}
	\langle k_{A'} | \hat{\boldsymbol{T}}_{j,-} | k_A \rangle
	   &= \langle k', l', m_{l'} | \boldsymbol{p}_{j,-} \hat U_j + \hat U_j \boldsymbol{p}_{j,-} | k, l, m_{l} \rangle
\end{align}
    The momentum eigenstates \(|\mathbf P\rangle\) (\(|\mathbf P'\rangle\)) are inserted between \(\hat U_j\) and \(\boldsymbol{p}_{j,-}\), with coefficients \(C_\alpha\) and \(C_\beta\) introduced for notation:
\begin{align}
	\langle k', l', m_{l'} | \hat{\boldsymbol{T}}_{j,-} | k, l, m_{l} \rangle
	   &= \sum_{\mathbf P} C_\alpha^* \langle \mathbf P | \hat U_j | k, l, m_{l} \rangle  \nonumber\\
	   &\quad + \sum_{\mathbf P'} C_\beta \langle k', l', m_{l'} | \hat U_j | \mathbf P' \rangle
\end{align}
    where \(C_\alpha^* = \langle k', l', m_{l'} | \boldsymbol{p}_{j,-} | \mathbf P \rangle\), \(C_\beta = \langle \mathbf P | \boldsymbol{p}_{j,-} | k, l, m_{l} \rangle\).\\
    For meson systems, the relations \(\boldsymbol{p}_{1,\pm} =\boldsymbol{p}_{\pm}\), \(\boldsymbol{p}_{2,\pm}=-\boldsymbol{p}_{\pm}\), \(\boldsymbol{r}_{1}=\dfrac{m_2}{m_1+m_2}\boldsymbol{r}\) and \(\boldsymbol{r}_{2}=-\dfrac{m_1}{m_1+m_2}\boldsymbol{r}\) hold. The matrix elements of the ladder operators are then evaluated explicitly. 
\begin{align}
	\boldsymbol{p}_\pm 
	   &= |\boldsymbol{p}|\,\hat{p}_\pm = |\boldsymbol{p}|\sin\theta e^{\pm i\phi}
	   = \mp \sqrt{\frac{8\pi}{3}} |\boldsymbol{p}| Y_{1,\pm 1}(\hat{\boldsymbol{p}})
\end{align}
    where \(Y_{1,\pm1}(\hat p)\) denote the rank-1 spherical harmonics with orbital angular momentum \(l=1\) and magnetic quantum numbers \(m=\pm1\), corresponding to the angular ladder components of momentum. 
    
    When the ladder operator \(\boldsymbol{p}_{\pm}\) acts on the initial state \(|k, l, m_{l}\rangle\), we obtain:
\begin{equation}
	\langle \mathbf P \vert \boldsymbol{p}_{\pm} \vert k, l, m_{l} \rangle
	   = \int d^3\mathbf P'' \langle \mathbf P \vert \boldsymbol{p}_{\pm} \vert \mathbf P'' \rangle
	   \langle \mathbf P'' \vert k, l,  m_l \rangle
\end{equation}
    For the matrix element \(\langle \mathbf P \vert \boldsymbol{p}_{\pm} \vert \mathbf P'' \rangle\), \(\boldsymbol{p}_{\pm}\) acts as the angular ladder operator of momentum, and its explicit form in momentum representation reads:
\begin{align}
	\langle \mathbf P \vert \boldsymbol{p}_{\pm} \vert \mathbf P'' \rangle
	   &= \mp \sqrt{\frac{8\pi}{3}} |\boldsymbol{p}| Y_{1,\pm 1}(\hat{\boldsymbol{p}})\, \delta^3(\mathbf P-\mathbf P'')
	\label{eq:p}
\end{align}
    Returning to the remaining factor in Eq.~\eqref{eq:p}, it should be noted that \(|k, l, m_{l}\rangle\) represents the wavefunction in coordinate representation, which needs to be transformed into momentum space.
    
    The corresponding momentum-space wavefunction is expressed as \(\psi_{k, l, m_{l}}(\boldsymbol{p})=\langle \mathbf P | k, l, m_{l}\rangle\).
    Using specific values of orbital quantum numbers l and \(m_l\), we can obtain the explicit expression for the matrix element \(\langle \mathbf P | \boldsymbol{p}_{\pm} | k,l,m_l\rangle\).
\begin{align}
	\langle \mathbf P | \boldsymbol{p}_{\pm} | 0\,0\,0 \rangle 
	   &= \mp i \sum_{n} C_{n} \big(2 \nu_{n}\big)^{\frac12} \psi_{0,1,\pm1}(\boldsymbol{p})  \\
	\langle \mathbf P | \boldsymbol{p}_{\pm} | 0\,1\,0 \rangle 
	   &= \mp i \sum_{n} C_{n} \big(2 \nu_{n}\big)^{\frac12} \psi_{0,2,\pm1}(\boldsymbol{p})  \\
	\langle \mathbf P | \boldsymbol{p}_{+} | 0\,1\,1 \rangle 
	   &= -\sqrt{2}\,i \sum_{n} C_{n} \big(2 \nu_{n}\big)^{\frac12} \psi_{0,2,2}(\boldsymbol{p})  \\
	\langle \mathbf P | \boldsymbol{p}_{-} | 0\,1\,1 \rangle
	   &= \frac{1}{\sqrt{3}} i \sum_{n} C_{n} \left(2 \nu_{n_i}\right)^{\frac12} \psi_{0,2,0}(\boldsymbol{p})  \nonumber\\
	   &\quad + i \sum_{n} C_{n} \big(2 \nu_{n}\big)^{\frac12} \psi_{0,0,0}(\boldsymbol{p})  \nonumber\\
	   &\quad + \sqrt{\frac{2}{3}} i \sum_{n} C_{n} \left(2 \nu_{n}\right)^{\frac12} \psi_{1,0,0}(\boldsymbol{p})
\end{align}

\end{document}